\documentclass[reprint,amsmath,amssymb,aps,prl,superscriptaddress]{revtex4-1}

\usepackage{times,color,amsmath}
\usepackage{subfigure,epsfig,graphicx,epstopdf}
\usepackage[none]{hyphenat}
\usepackage[colorlinks]{hyperref}
\hypersetup{linkcolor = {blue},	citecolor = {blue},	urlcolor = {blue}}

\newcommand{\affa}{\affiliation{Department of Applied Physics, KTH Royal Institute of Technology, Albanova University Centre, Roslagstullsbacken 21, 106 91 Stockholm, Sweden}}
\newcommand{\affb}{\affiliation{Nordita, Stockholm University and KTH Royal Institute of Technology, Hannes Alfvéns väg 12, SE-106 91 Stockholm, Sweden}}
\newcommand{\affc}{\affiliation{Institute for Physics of Microstructures, Russian Academy of Sciences, 603950 Nizhny Novgorod, GSP-105, Russia}}

\begin{document}

\title{ Observation of reentrant metal-insulator transition in a random-dimer disordered SSH lattice}

\author{Ze-Sheng Xu} \email{zesheng@kth.se} \affa     
\author{Jun Gao}  \email{junga@kth.se} \affa 
\author{Adrian Iovan} \affa 
\author{Ivan M. Khaymovich} \affb \affc
\author{Val Zwiller} \affa 	
\author{Ali W. Elshaari} \email{elshaari@kth.se} \affa 

\date{\today}

\begin{abstract}

The interrelationship between localization, quantum transport, and disorder has remained a fascinating focus in scientific research. Traditionally, it has been widely accepted in the physics community that in one-dimensional systems, as disorder increases, localization intensifies, triggering a metal-insulator transition. However, a recent theoretical investigation [Phys. Rev. Lett. 126, 106803] has revealed that the interplay between dimerization and disorder leads to a reentrant localization transition, constituting a remarkable theoretical advancement in the field. Here, we present the experimental observation of reentrant localization using an experimentally friendly model, a photonic SSH lattice with random-dimer disorder, achieved by incrementally adjusting synthetic potentials. In the presence of correlated on-site potentials, certain eigenstates exhibit extended behavior following the localization transition as the disorder continues to increase. We directly probe the wave function in disordered lattices by exciting specific lattice sites and recording the light distribution. This reentrant phenomenon is further verified by observing an anomalous peak in the normalized participation ratio. Our study enriches the understanding of transport in disordered mediums and accentuates the substantial potential of integrated photonics for the simulation of intricate condensed matter physics phenomena.

\end{abstract}

\maketitle

Since the pioneering work of Rolf Landauer and Philip. W. Anderson\cite{La, And}, the phenomenon of wave function localization in disordered media has been extensively investigated, exploring the quantum phase transition and the transport physics in disordered systems. Anderson localization predicts the absence of diffusive behavior in the single-particle wave function when subjected to a disorder potential, arising from wave interference effects between multiple propagation paths. As the disorder strength increases, transport becomes suppressed resulting in the metal-insulator transition. Extensive studies have been conducted on this transition in diverse systems, including electron gases, doped semiconductors, random media, photonic lattices, ultra-cold optical lattices, ultrasound, and atomic system~\cite{Cut, PAL, Diederik, Martin, Tal, Topol, Hefei, AA, Donald}. The prior work on low-dimensional systems ($D \leq 2$) establishes that uncorrelated disorder can shift all eigenstates to a localized phase, regardless of the disorder degree~\cite{EAbr}. In contrast, a quasiperiodic potential reveals that a non-zero critical disorder strength can differentiate the metallic and insulator phases, generating an intermediate critical region where both extended and localized states coexist~\cite{SD, HP, SA, YW, MG, SG, beta-footnote}. This critical region is defined by the single-particle mobility edge~\cite{PAL, NMO, Biddle2009, XLI, SD2, SD3, MGR, JBI}, symbolizing a critical energy threshold distinguishing the boundary between the extended and localized states in the system.

Historically, the prevailing belief held that the system would undergo a unidirectional transition from the extended state to the localized state as the disorder increases. Contrary to this understanding, recent theoretical studies illustrate that the reentrant localization can be achieved in a Su-Schrieffer-Heeger (SSH) chain with staggered quasiperiodic disorder~\cite{SRO1, SRO2} or random-dimer disorder~\cite{ZWZ}. This implies that with a further increase in disorder, two transitions ensue, leading to an extended phase with the mobility edge and further to the second localized phase, respectively. The genesis of this phenomenon resides in the increase of the disorder strength, which results in the re-extension of some previously localized states. At even higher disorder strength these states converge to the localized phase, showing a strong correlation with multiple critical regions that accommodate the mobility edge within the spectrum.

In this work, we implement a correlated disorder by imposing a specific potential on the random dimer of an SSH chain. To achieve this, we employ a Si$_{3}$N$_{4}$ waveguide array fabricated using electron beam lithography and reactive ion etching~\cite{JG1, ZSX, JG2}, facilitating effective control over the nearest-neighbor hopping amplitude. In addition, we tailor the width of array waveguides to introduce a synthetic on-site potential~\cite{JG3, JG4}. To examine the system, we selectively excite a specific single site within the waveguide array. Subsequently, we adopt a top imaging strategy to probe the intensity distribution at the end of the array by analyzing the scattered light. These images are recorded for different disorder strengths, and the normalized participation ratios (NPR), characterizing the localization-delocalization phase diagram, are extracted. In addition to observing light diffusion or localized behavior corresponding to different quantum transport phases, we also discern a consistent trend in the NPR of single-site excitation that aligns with our theoretical model. The concordance between experimental observations and theoretical predictions provides compelling evidence of the non-monotonic metal-insulator transition in this one-dimensional photonic system. This reentrant localization transition, observed experimentally, not only opens a promising pathway for studying condensed matter physics within an on-chip nano-photonic system but also broadens the frontiers of quantum encoding and transport in low-dimensional systems, with potential applications in quantum information processing and communication~\cite{FUL, DAN, ANDR, JIA}.

Here, we consider a one-dimensional SSH model with random-dimer disorder, which is described by the following Hamiltonian

\begin{equation}\label{eq:1}
  \begin{aligned}
   H=t \sum_{n=1}^{N-1}\left[1 - (-1)^{n} \Delta](c_n^{\dagger} c_{n+1}+\text { H.c. }\right) \\ +  \sum_{n=1}^{N/2} \epsilon_n (c_{2n-1}^{\dagger} c_{2n-1}+c_{2n}^{\dagger} c_{2n}),
   \end{aligned}
\end{equation}

\begin{equation} \label{eq:2}
\epsilon_n= \begin{cases}\epsilon, &\text{ probability} = p  
\\ 0, & \text { probability } = 1-p\end{cases}
\end{equation}

This model comprises $N/2$ unit cells, each containing two sites, where $c_n^{\dagger}$ and $c_n$ represent the creation and annihilation operators corresponding to the site $n$. The nearest-neighbor hopping $t(1\pm \Delta)$, with the dimerization parameter $\Delta$, represent the intra-cell and inter-cell coupling strength between sites, respectively. $\epsilon_n$ designates the bimodally distributed on-site potential applied to the random dimer, being $\epsilon$ with the probability $p$ and zero otherwise. By tuning the value of $\epsilon$, we can probe the quantum phase transitions occurring at different disorder strengths. This model presents the reentrant localization at the specific critical values of $\epsilon$, which can be examined by analyzing the $\mathrm{NPR}$  in conjunction with the inverse participation ratio ($\mathrm{IPR}$ )~\cite{XLI, XLI2}. The definitions of these metrics are provided as follows:

\begin{equation} \label{eq:3}
\mathrm{NPR}_n=\left(N \sum_{i=1}^N\left|\phi_n^i\right|^4\right)^{-1}, \quad \mathrm{IPR}_n=  \sum_{i=1}^N\left|\phi_n^i\right|^4. 
\end{equation}

Where $\phi_n^i$ denotes the $n$-th eigenstate coefficient at the site $1\leq i\leq N$, while $N$ signifies the number of sites in the lattice. It is theoretically proposed that in the thermodynamic limit, as $N$ tends towards infinity, the $\mathrm{NPR}$ stays finite when the $\mathrm{IPR}$ vanishes for an ergodically extended state. Conversely, the $\mathrm{NPR}$ approaches zero when the $\mathrm{IPR}$ stays finite for a localized state. 
Further, we will use the mean values across the spectrum of the above two measures, defined as follows

\begin{equation} \label{eq:4}
\langle\mathrm{NPR}\rangle=\frac{1}{N} \sum_{i=1}^N \mathrm{NPR}_n  , \quad \langle\mathrm{IPR}\rangle=\frac{1}{N} \sum_{i=1}^N \mathrm{IPR}_n .
\end{equation}

\begin{figure}[!t]
	\centering
	\includegraphics[width=1.0\linewidth]{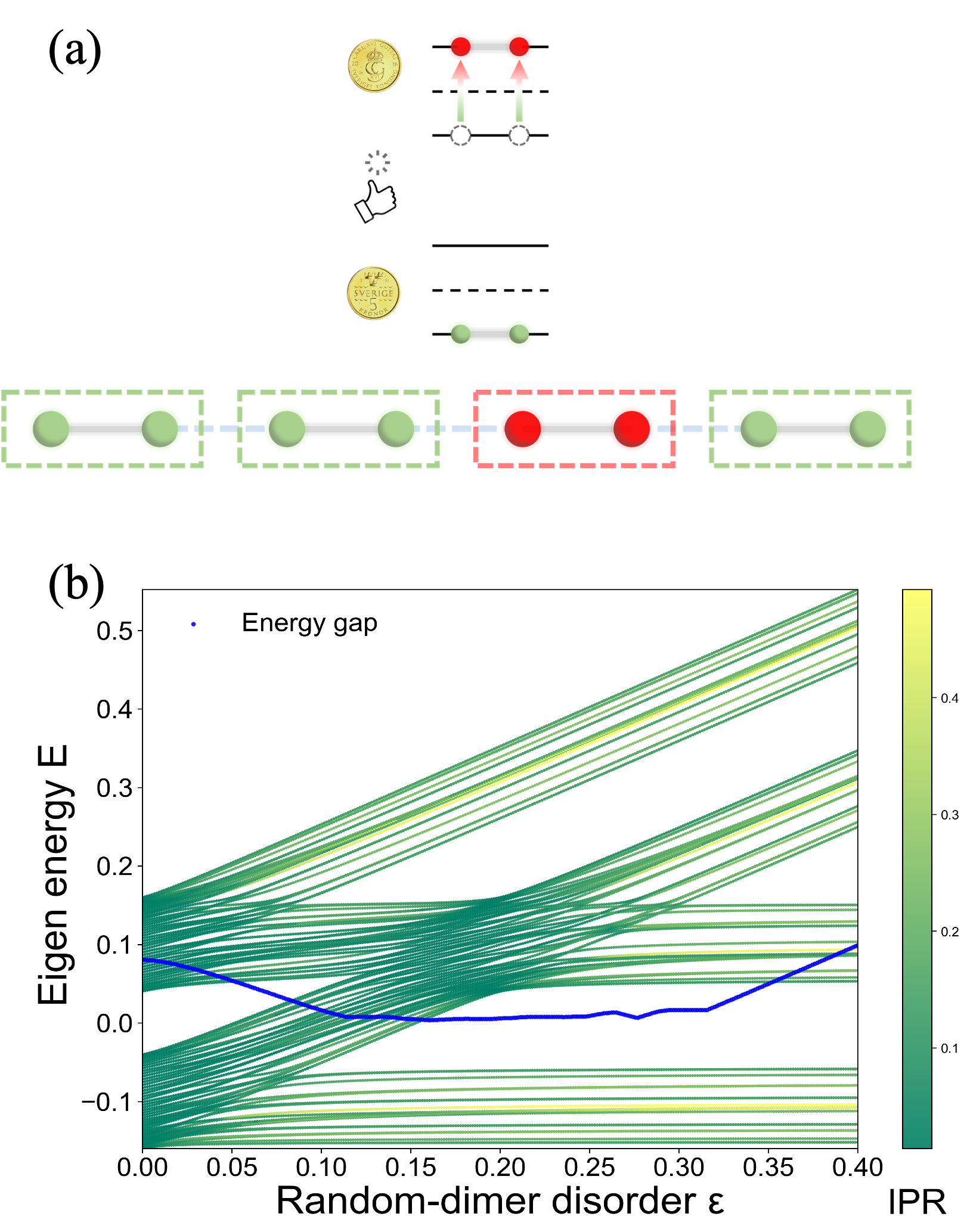}
	\caption{\textbf{SSH lattice with random-dimer disorder and dimerization amplitude $\Delta$ = 0.25 featuring an on-site potential.} \textbf{(a)}~Schematic representation of the random-dimer disordered SSH model. The two green nodes within the dotted box depict a situation devoid of on-site potential, whereas the two red nodes symbolize a dimer with a chosen finite on-site potential. The on-site potential is randomly allocated with a probability $0.5$, analogous to the likelihood of getting heads in a fair coin toss. In a broad sense, our model's on-site potential can be regarded as a bimodal telegraph noise.  
    \textbf{(b)}~Eigenenergy spectrum of the random-dimer disordered SSH model with varying on-site potential $\varepsilon$. The spectrum is clearly separated into four bands: energies of two of them increase linearly with $\epsilon$, while those of the other two are insensitive to it. The corresponding wave functions of linear (constant) in $\epsilon$ live mostly at disordered $\epsilon_n=\epsilon$ (clean $\epsilon_n=0$) dimers. In each pair of bands, these wave functions are also symmetric for the lower and antisymmetric for the higher one of two bands, see \textbf{Supplemental Information}.
 As the disorder strength approaches two critical values, we observe the gap closure and reopening, indicative of the disappearing and reappearing in-gap edge states, associated with the non-monotonic insulator-metal-insulator transition. The blue curve represents the Energy gap in this system.}
	\label{f1}
\end{figure}

\begin{figure*}[htbp]
	\centering
	\includegraphics[width=0.95\linewidth]{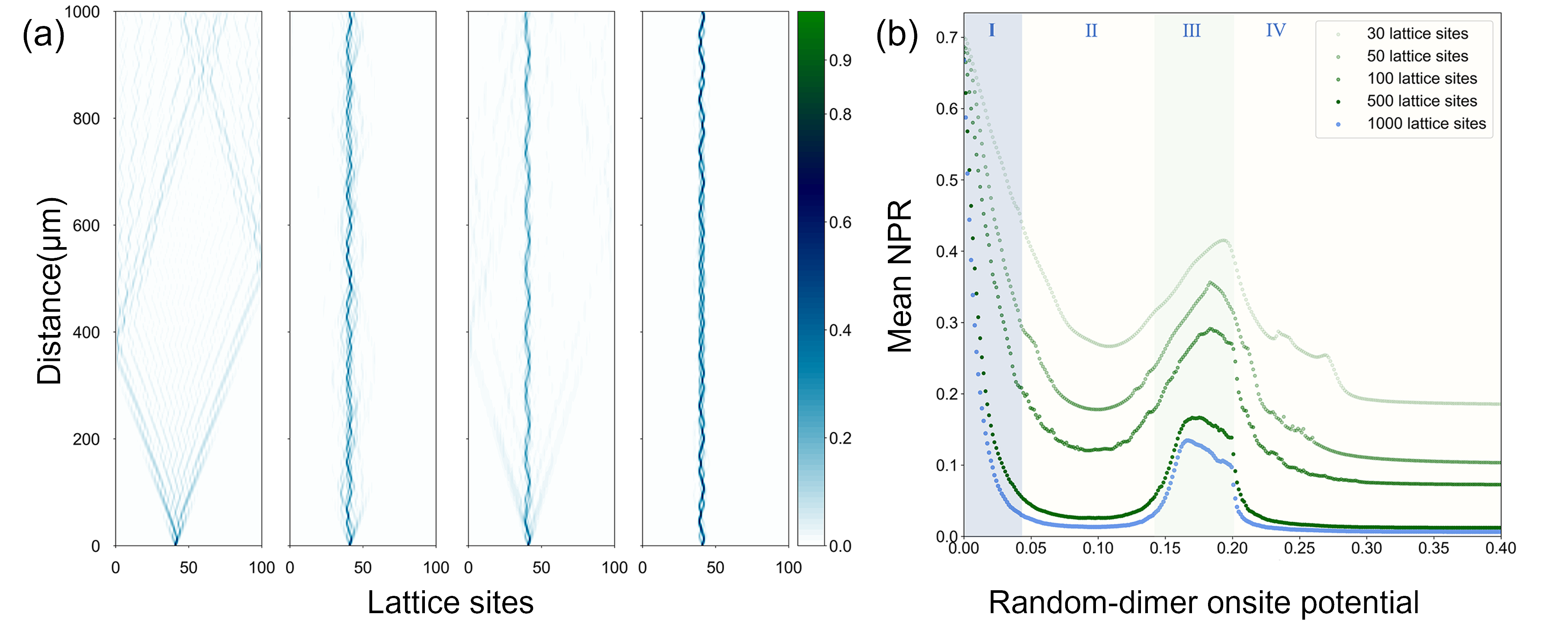}
	\caption{\textbf{Light transport in the lattice and average NPR under varying random-dimer on-site potential.} \textbf{(a)}~The four simulated trajectories of light in $1000~\mu$m long SSH lattices comprising $100$ sites with varying degrees of random dimer disorder are displayed.  \textbf{I}, \textbf{II}, \textbf{III}, and \textbf{IV} correspond to random-dimer on-site potentials, $\varepsilon = 0$, $0.1$, $0.185$, and $0.3$ respectively, with random-dimer assignment probability of $p=0.5$. In \textbf{I}, a ballistic light propagation takes place without any disorder present; in \textbf{II} and \textbf{IV}, light becomes localized within the lattice at the designed random-dimer on-site potentials, despite significant propagation distance. \textbf{III} demonstrates the phenomenon of reentrant localization, where both extended and localized states coexist. \textbf{(b)}~The numerically computed mean NPR as a function of random-dimer potential $\epsilon$ over a range of lattice sites from $30$ to $1000$. As the number of lattice sites increases, the mean NPR exhibits a declining trend, yet the counterintuitive reentrant peaks consistently emerge. The four light transport simulations coincide with the different regimes in the mean NPR plot, as denoted by the color coding and labeling \textbf{I} to \textbf{IV}.}.
	\label{f2}
\end{figure*}

When both the average values $\langle \mathrm{NPR} \rangle$ and $\langle \mathrm{IPR} \rangle$ are simultaneously finite, it suggests the coexistence of localized and extended states. This gives rise to intermediate states, triggering the phenomenon of reentrant localization and inducing the non-monotonic metal-insulator transition within the system. However, relying solely on the $\mathrm{NPR}$ or $\mathrm{IPR}$ values in our investigation poses difficulties in accurately identifying the coexistence of eigenstates within the extended and localized phases. One feasible approach is the simultaneous calculation of the $\mathrm{NPR}$ and $\mathrm{IPR}$ along the random-dimer onsite potential, followed by the analysis of the non-zero region for both these parameters. An alternative strategy involves computing a defined parameter $\eta$ as $\eta=\log _{10}[\langle\mathrm{IPR}\rangle \times\langle\mathrm{NPR}\rangle]$~\cite{XLI2}. By assessing the ratio of extended to localized states within all N eigenstates, we establish the criteria for the reentrant regime. The detailed derivation can be found in the \textbf{Supplementary information}:

\begin{equation} \label{eq:5}
\mathrm{log_{10}}\left(\frac{2}{N}\right) \leq \eta \leq \mathrm{log_{10}}\left(\frac{1}{4}+\frac{1}{2N}\right)
\end{equation}

In our experiment, we select a lattice with $N = 100$ sites, as depicted in Fig.~\ref{f1}(a), and generate a random-dimer number array with a probability $p = 0.5$ according to Eq. (\ref{eq:2}). We choose $t = 0.08$ and $\Delta = 0.25$, so the intra/inter-cell coupling strength is $0.1$ and $0.06$, which sets the lattice at the topologically trivial phase~\cite{ZWZ}. Fig.~\ref{f1}(b) presents the results of sweeping the on-site potential from $0$ to $0.4$ across the random-dimer number array. Subsequently, we conduct numerical calculations to determine the lattice's spectrum, unveiling the system's energy levels and their corresponding IPR. 
The spectrum of the system is given by four well distinguished bands, centered around $E_{\nu,\pm}=\nu\epsilon \pm t(1+\Delta)$, with wave functions, living mostly on the dimers with the on-site potential $\epsilon_n =\nu\epsilon$, $\nu=0,1$, and being symmetric (antisymmetric) at each of such dimers with the corresponding energy shift $\mp t(1+\Delta)$, respectively, see \textbf{Supplemental Information}. The width of each band is given by the smaller hopping $2t(1-\Delta)$. The blue curve uncovers the closing and reopening of the energy gap in the system, signifying the topological phase transition of the eigenstates. It is this gap closure which is responsible for the reentrant delocalization. Indeed, close to the intersection of $2$ bands, the effective disorder is given by the difference of energy centers $E_{\nu,\pm}$. As soon as $2$ bands cross, this difference goes to zero and a disorder-free sector of the Hamiltonian with delocalized states emerges at this crossing-band pair. The states in the other pair of bands stay localized. This explains the reentrant character of delocalization-localization phase diagram. 

Fig.~\ref{f2}(a) displays the simulated pattern of the $41$st single-site excitation obtained using the coupled mode theory~\cite{SEM, JRP, HAH} to showcase the evolution of light within lattices having a length of $1000~\mu$m. The on-site potentials $\epsilon$ of these  lattices are $0$, $0.1$, $0.185$ and $0.3$, from left to right. Comparing  Fig.~\ref{f2}(a) \textbf{III} with \textbf{II} and \textbf{I}, we observe that the light undergoes a distinct behavior in \textbf{III}. In this intermediate regime \textbf{III}, a part of the light intensity stays near the initial location as in \textbf{II}, but another part ballistically spreads as in \textbf{I1}. This gives evidence of a coexistence of localized and extended light, observed simultaneously. Unexpectedly this coexistence appears even after the light has transitioned into the fully localized regime \textbf{II} upon reaching the critical value of the on-site potential, $\epsilon$. Furthermore, as we continuously increase $\epsilon$, the light evolution pattern in regime \textbf{IV} reverts back to exhibiting localized behavior, clearly demonstrating the occurrence of reentrant localization in this system. Fig.~\ref{f2}(b) shows the numerically calculated Mean NPR for the random-dimer on-site potential system with pre-defined parameters above while varying lattice sites numbers.  In the limit of a large number of lattice sites, the $\langle \mathrm{NPR} \rangle$ approaches its lower bound, which scales as $\sim \frac{1}{\mathrm{N}}$. Despite the decreasing trend of the lower limits as the number of sites increases, the curves consistently exhibit an anomalous peak followed by a subsequent decrease. This observed behavior of the curves provides convincing evidence for the existence of reentrant localization in this system and justifies that the reentrant localization in our system with $100$ sites does exist and can be experimentally observed.

\begin{figure}[!t]
	\centering
	\includegraphics[width=0.95\linewidth]{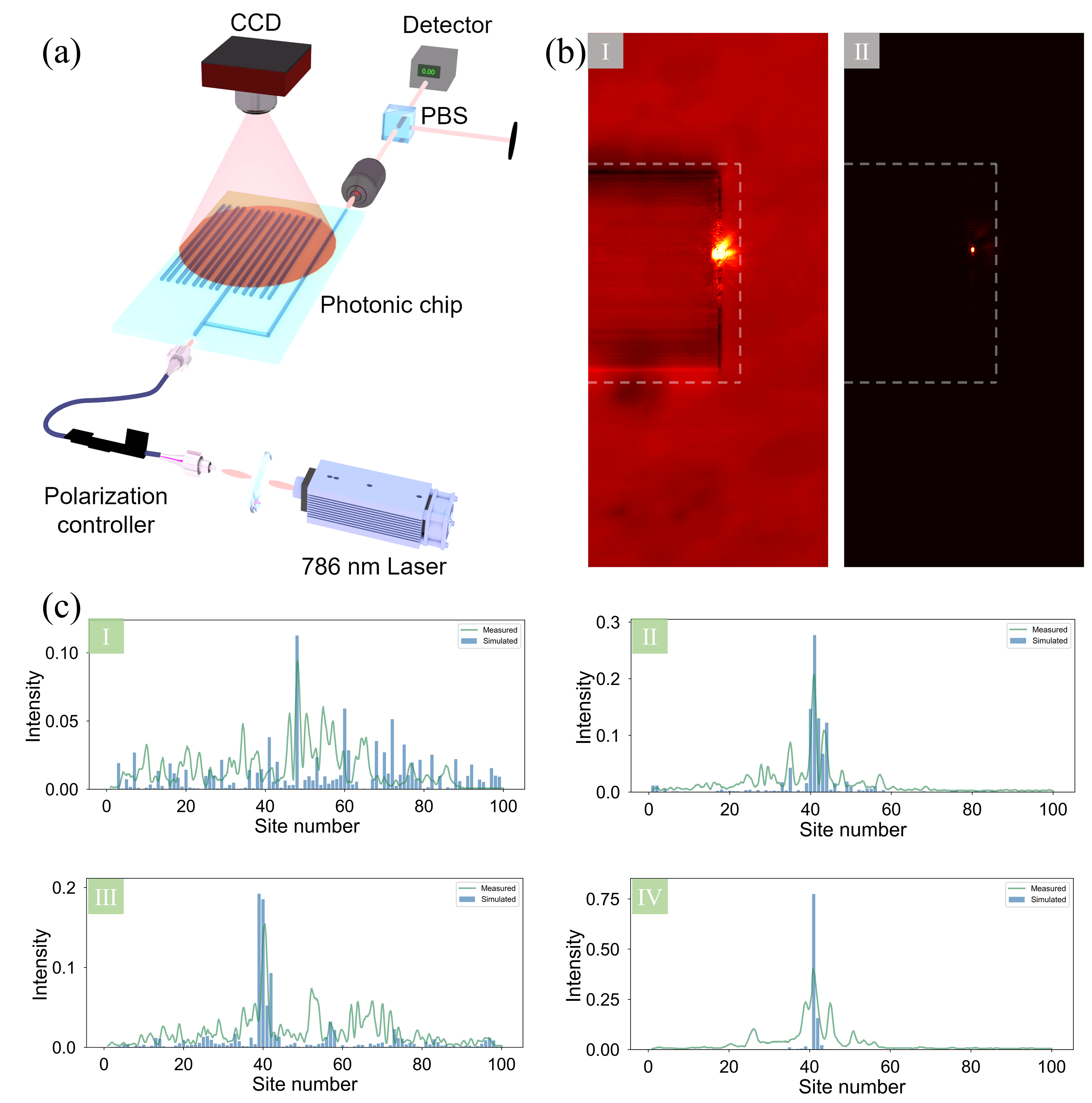}
	\caption{\textbf{Details of the experimental setup and collected data.} \textbf{(a)}~Illustration of the experimental setup. A $786$~nm continuous wave laser is utilized for the excitation of the random-dimer disordered SSH photonic lattice through a lensed fiber. A Y-splitter is incorporated to control polarization and optimize the coupling to the photonic chip. The Transverse Electric (TE) mode is excited by accurately adjusting the paddle of the polarization controller and monitoring the detector counts transmitted through the chip. A polarizing beam splitter enables exciting the TE modes by minimizing the detector count. The intensity of the facet light distribution is imaged using a microscope objective and a CCD camera. \textbf{(b)}~Extraction of the output light distribution. Two images, both with identical focus, are captured to ascertain the distribution of light intensity. \textbf{II} represents the output pattern of the lattice, while \textbf{I} serves as a reference pattern of the illuminated lattice that aids in pinpointing exact coordinates in pixel space. The section within the dotted line in both images indicates the position of the lattice, obtained from the reference image, to extract the light intensity at each lattice site. \textbf{(c)}~Comparison between experimentally measured and simulated light intensity distributions. \textbf{I}, \textbf{II}, \textbf{III}, and \textbf{IV} correspond to the output intensity distribution at a distance of $1000~\mu$m, following the $41$st single-site excitation with random-dimer on-site potentials $\varepsilon = 0$, $0.1$, $0.185$, and $0.3$, respectively.}
	\label{f3}
\end{figure}

\begin{figure*}[htbp]
	\centering
	\includegraphics[width=1.0\linewidth]{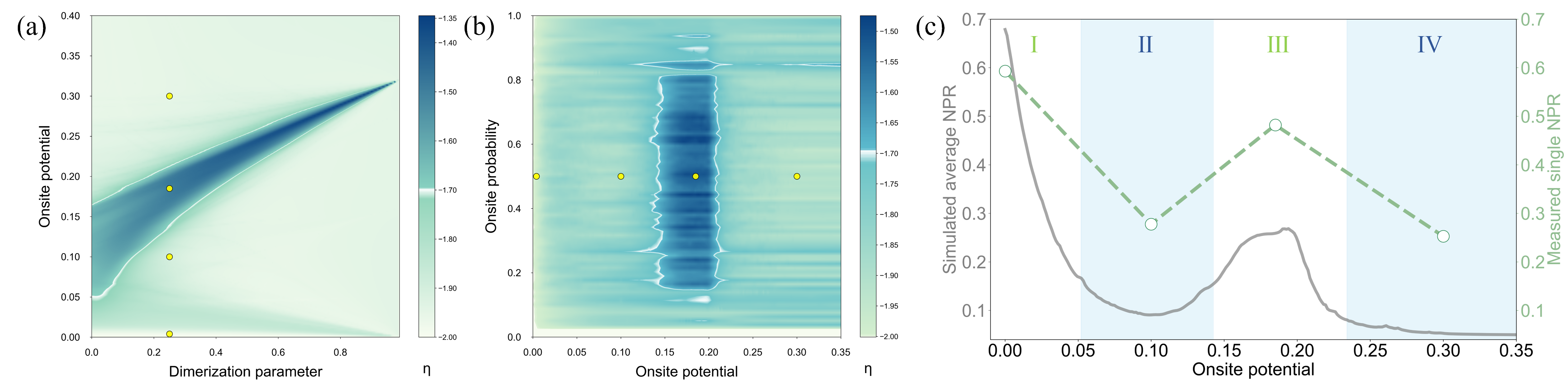}
	\caption{\textbf{Phase diagram and retrieved NPR for single-site excitation.} \textbf{(a)}~The phase diagram of $\eta$ over the dimerization parameter $\Delta$ and on-site potential $\epsilon$ plane for a $100$-site dimer-disordered SSH lattice. The blue region signifies the concurrent non-zero values of both $\langle\mathrm{IPR}\rangle$ and $\langle \mathrm{NPR} \rangle$, according to the definition $\eta=\log _{10}[\langle\mathrm{IPR}\rangle \times\langle\mathrm{NPR}\rangle]$. Four small yellow circles in the diagram represent the experimental sampling points we selected, corresponding to the four regions presented in Fig.~\ref{f2}(b). The reentrant localization is validated by the coexistence of localized and extended states, a consequence of the system hosting a mobility edge. \textbf{(b)}~Phase diagram of $\eta$ in the on-site potential $\epsilon$ and on-site probability $\mathrm{p}$ plane. The near-vertical border of the reentrant localization region accentuates that the critical values remain stable as the on-site probability varies widely ($0.15<q<0.8$), confirming that our choice of on-site probability $p=0.5$ does not add any specificity. \textbf{(c)}~The trend of the average NPR results from the ensemble of statistical behaviors of all NPRs from single-site excitations. Specifically, the NPRs from single-site excitations which are linked with extended eigenstates contribute to the peak of the average NPR in the reentrant localization and trigger a non-monotonic metal-insulator transition in the one-dimensional random-dimer disordered SSH system.}
		\label{f4}
\end{figure*}

To investigate our system's properties, we employ a full-vectorial mode solver~\cite{FDTD} for numerical simulations of the waveguide's propagation constants. These propagation constants serve as a representation of the on-site potential within our model. Detailed analysis is presented in the \textbf{Supplementary information}. In our experiment, we utilized a CMOS-compatible Si$_{3}$N$_{4}$ photonic platform to successfully integrate nano-sized SSH devices with random-dimer disorder. Each device consists of $N=100$ waveguides, with dimensions of $250$~nm high and $450-518$~nm wide according to the different on-site potentials, with SiO$_{2}$ bottom and air top cladding. These waveguides support only the fundamental modes at a $786$~nm wavelength. The width of and the gap between these waveguides are well-engineered to satisfy the designed hopping amplitude and the on-site potentials. The experimental setup is shown in Fig.~\ref{f3}(a). We excite the lattice with a continuous-wave coherent laser centered at $786$~nm. Using a lensed fiber fixed on a six-axis nano-positioning stage, the laser is finely coupled into the nanophotonic chip. A polarizing beam splitter with a three-paddle polarization controller is installed to selectively excite only the transverse electric (TE) mode. The on-chip Y-splitter divides the light into two paths, the first leading to the random-disordered SSH lattice, with the second acting as the monitor waveguide for tuning the coupling efficiency and controlling the polarization. The output pattern of the light intensity at the end of each lattice is collected from the scattered signal imaged with a $40$X objective and captured by a charge-coupled device (CCD) camera. Fig.~\ref{f3}(b) presents captured images obtained under both non-illuminated and illuminated conditions, with the focus remaining consistent. By analyzing the region marked by dashed lines in Fig.~\ref{f3}(b) \textbf{I} , we extract the pixel coordinates corresponding to the photonic lattice structure. By combining this information with Fig.~\ref{f3}(b) \textbf{II} which depicts the output pattern of the lattice, we establish a relationship between the light intensity distribution and the lattice site numbers ranging from $1$ to $100$. This enables us to retrieve the $\mathrm{NPR}$ for the single-site excitation scenario, providing valuable insights into the system's light behavior. Fig.~\ref{f3}(c) shows the experimentally measured light distribution along the sites compared with the theoretically predicted light distribution. We can clearly see the distinct signature of extended and localized wave packets transport in different regimes from \textbf{I} to \textbf{IV}. \textbf{I} indicates extended behavior with $\epsilon = 0$, while \textbf{II} and \textbf{IV} demonstrate localized behavior. \textbf{III} represents the intermediate case. Specifically, in region \textbf{II} and \textbf{IV} the wave packet is highly confined within few lattice sites, while in region \textbf{III} we see a clear signature that indicates the coexistence of extended and localized behaviour. The peak and overall behavior of the measured light intensity distribution align well with the theoretical prediction. Notably, the laser light transport shows consistency with the eigenstates' behavior, implying the non-monotonic metal-insulator transition and the reentrant localization.

In order to explore the reentrant localization regime, we plot the phase diagram according to Eq.~\ref{eq:5}. Fig.~\ref{f4}(a) reveals the $\eta$ over different dimerization parameters $\Delta$ and on-site potentials $\mathrm{\epsilon}$, while Fig.~\ref{f4}(b) shows the $\eta$ for different on-site potentials $\mathrm{\epsilon}$ and on-site probabilities $\mathrm{p}$. The white edges in these phase diagrams separate the localized phases from the delocalized one with the mobility edge. Non-monotonic behavior with the on-site potential shows the reentrant localization which we are interested in. The four yellow circles mark the experimental sampling points for collecting the light distribution and calculating the single-site excitation $\mathrm{NPR}$. Fig.~\ref{f4}(c) presents the $\mathrm{NPR}$ of single-site excitation compared with the simulated $\langle \mathrm{NPR} \rangle$. Four highlighted regions \textbf{I}, \textbf{II}, \textbf{III}, \textbf{IV} correspond to the four specific regimes depicted in Fig.~\ref{f2}, respectively. The measured $\mathrm{NPR}$ from the $41$st single-site excitation contributes to the statistical behavior observed in the averaged $\langle \mathrm{NPR} \rangle$ and exhibits a consistent trend, as depicted in the plot. The choice of the $41$st site to be excited is not specific and for most of the sites the behavior is expected to be similar. The $\mathrm{NPR}$ function shows a non-monotonic trend, it decreases in value as the disorder strength increases, subsequently revealing an anomalous peak in the intermediate regime, and ultimately undergoing a secondary decrease for large disorder strength. Based on the experimental results, we unambiguously observe the occurrence of reentrant localization behavior in the random-dimer disordered SSH photonic lattices, both through intuitive visual inspection and rigorous multifractal analysis.

In conclusion, we have demonstrated the observation of the reentrant localization in a one-dimensional disordered system. This system is based on a nano-fabricated Si$_{3}$N$_{4}$ platform realized in photonic waveguide devices. The numerically calculated $\langle \mathrm{NPR} \rangle$ and the phase diagram for the defined parameter $\eta$ guided our choice of sampling points with different on-site potentials $\epsilon$ to illustrate the transitions. Our technique of fine-tuning the width of waveguides allows us to control the on-site potential for each dimer. By using a single-shot approach, which involves probing the light intensity at each lattice site using an edge-scattered structure, we can investigate and observe the transport of light within the system. This approach enables us to retrieve the normalized participation ratio ($\mathrm{NPR}$) for the single-site excitation scenario. The counter-intuitive drop and increase of $\mathrm{NPR}$ indicate the occurrence of non-monotonic localization transitions, thus pointing to the presence of reentrant localization. Our work contributes to a deeper comprehension of the underlying mechanisms governing the behavior of quantum systems subjected to disorder. Such insights are invaluable in elucidating the complex nature of localization transitions and their implications in various physical systems. From a technical perspective, we incorporate a nanometer-level precision electron beam lithography process, resulting in a  precise implementation of complex on-site potentials and hopping parameters. The field of micro-nano photonics, particularly with the inclusion of the reentrant localization, holds a significant potential in various areas such as information processing, optical computing, hybrid photonic circuits, topological photonics, and photonic quantum technologies~\cite{AWE1, BAE, TOZ, LLU, CPO}.

\begin{acknowledgments}
A.W.E acknowledges support Knut and Alice Wallenberg (KAW) Foundation through the Wallenberg Centre for Quantum Technology (WACQT), Swedish Research Council (VR) Starting Grant (Ref: 2016-03905), and Vinnova quantum kick-start project 2021. V.Z. acknowledges support from the KAW and VR.
Work at Nordita was supported by  European Research Council under the European Union Seventh Framework  ERS-2018-SYG HERO, KAW 2019.0068 and the University of Connecticut.
\end{acknowledgments}


\begin{thebibliography}{99}
\bibitem{beta-footnote} In some short-range uncorrelated disordered models, the coexistence of extended and localized states also appears and do not show the mobility edge formation, see, e.g.,~\cite{beta-ens}.
\end{thebibliography}
\end{document}